\documentclass[a4paper,11pt]{article}
\usepackage{pos}
\newcommand{\sparse}{\mathcal{S}}

\title{Nucleon-Pion Spectroscopy from Sparsened Correlators}
\author*[a]{Anthony V. Grebe}
\author[a]{Michael Wagman}

\affiliation[a]{Fermi National Accelerator Laboratory,\\
  Kirk and Pine Street, Batavia, IL, USA}

\emailAdd{agrebe@fnal.gov}

\abstract{Neutrino oscillation experiments require accurate reconstructions of neutrino energies, which depend in part on a theoretical understanding of the axial $N \rightarrow \Delta$ transition form factors. Lattice QCD studies of this transition require construction of all hadronic states with energies up to the mass of the $\Delta$ resonance, which includes $N\pi$ and $N\pi\pi$ for physical quark mass values. Building interpolating operators from sparse grids of source and sink points is a versatile method of approximating all-to-all quark propagators that has been successfully used in other multi-hadron calculations. This work will discuss applications of this method to $N\pi$ and $N\pi\pi$ systems and present preliminary results.}

\FullConference{The 40th International Symposium on Lattice Field Theory (Lattice 2023)\\
July 31st - August 4th, 2023\\
Fermi National Accelerator Laboratory\\}

\begin{document}
\maketitle

\section{Introduction}

Maximizing the discovery potential of DUNE and other future accelerator neutrino experiments will require precise and accurate predictions for neutrino-nucleus cross sections~\cite{Acciarri:2015uup}.
Making cross-section predictions using controlled approximations to the Standard Model is essential for discovering beyond-the-Standard-Model (BSM) physics at neutrino experiments~\cite{Coyle:2022bwa}.
Lattice quantum chromodynamics (QCD) is needed to provide first principles determinations of the nucleon-level processes that underlie nuclear effective field theories (EFTs) and phenomenological models of neutrino-nucleus scattering~\cite{Kronfeld:2019nfb,Meyer:2022mix,Ruso:2022qes}.
Studies of electroweak pion production will be especially important for DUNE, where the energy flux will be broadly peaked around the 100 MeV to few GeV region where resonance production is common~\cite{Mosel:2019vhx,Rocco:2020jlx,Lovato:2023raf}.

The $\Delta(1232)$ plays an important role in these processes, and achieving percent-level precision in nuclear-many-body predictions of neutrino-nucleus scattering will require theoretical predictions of $N \rightarrow \Delta$ transition form factors with few-percent precision~\cite{Simons:2022ltq}.
Current phenomenological determinations and QCD model predictions of axial $N\rightarrow \Delta$ transition form factors only achieve 10-20\% precision~\cite{Hernandez:2007qq,Hernandez:2010bx, light-cone-rules-1, light-cone-rules-2}.
Lattice QCD studies of pion- and $\Delta$-production amplitudes that could improve constraints on these and other form factors are therefore essential for achieving percent-level precision on neutrino-nucleus cross sections.

Exploratory lattice QCD studies of $N \rightarrow \Delta$ transition form factors have been performed using heavier-than-physical quark masses for which the $\Delta$ is stable~\cite{Alexandrou:2006mc,Alexandrou:2010uk,Alexandrou:2013ata,Alexandrou:2015hxa}.
Extending these studies to physical quark masses, where the $\Delta(1232)$ is a resonance capable of decaying to one- and two-pion states, requires lattice QCD calculations that can disentangle resonant and non-resonant states in the vicinity of the $\Delta$ resonance.
This can be achieved by calculating correlation functions with a large set of $N$, $N\pi$, $N\pi\pi$, and $\Delta$ interpolating operators and then determining the spectrum and matrix elements of QCD energy eigenstates using variational methods~\cite{Fox:1981xz,Michael:1982gb,Luscher:1990ck,Blossier:2009kd}.
These finite-volume results can then be matched to infinite volume $N\rightarrow \Delta$ and non-resonant $N \rightarrow N\pi$ and $N \rightarrow N\pi\pi$ amplitudes through generalizations of the Lellouch-L{\"u}scher formula under active development~\cite{Lellouch:2000pv,Briceno:2014uqa,Briceno:2015csa,Baroni:2018iau,Muller:2020wjo,Hansen:2021ofl} or used to constrain finite-volume nuclear effective theories.
The same variational calculations will result in determinations of nucleon elastic axial form factors, another crucial input needed to reduce uncertainties in neutrino-nucleus cross sections~\cite{Meyer:2016oeg,Meyer:2022mix,Ruso:2022qes,Simons:2022ltq,MINERvA:2023avz}, with the contributions from $N\pi$ excited states explicitly removed.
This will provide important validation for other current and future methods of removing significant $N\pi$ excited-state contamination from lattice QCD studies of nucleon axial form factors~\cite{Bali:2018qus,RQCD:2019jai,Jang:2019vkm,Alexandrou:2020okk,Park:2021ypf,Djukanovic:2022wru}.

The first step in this process is the construction of a set of interpolating operators that can describe the QCD energy eigenstates with baryon number one and isospin $I = 1/2$ and $I = 3/2$ up through the $\Delta(1232)$ resonance.
The inclusion of $N\pi$ interpolating operators built from spatially nonlocal products of plane-wave $N$ and $\pi$ operators has been found to be essential for resolving the low-energy spectra in these channels~\cite{Lang:2012db,Kiratidis:2016hda,Lang:2016hnn,Andersen:2017una,Silvi:2021uya,Bulava:2022vpq,Alexandrou:2023elk}, and the first steps towards computing $N\rightarrow N\pi$ form factors using these operators have recently been reported in Ref.~\cite{Barca:2022uhi}.

This work will present calculations of $N$, $N\pi$, and $N\pi\pi$ correlation functions that permit the study of the non-resonant states in the vicinity of the $\Delta(1232)$ resonance.
It will describe an algorithm for calculating the computationally demanding $N\pi\pi \rightarrow N\pi\pi$ correlation functions using quark propagator sparsening~\cite{nplqcd-sparsening, xu-feng-sparsening, nplqcd-variational} and sequential pion propagators as well as the software implementation and optimizations used to enable efficient calculations on GPUs.

\section{Methodology}

\begin{figure}
    \centering
    \includegraphics[width=0.4\textwidth]{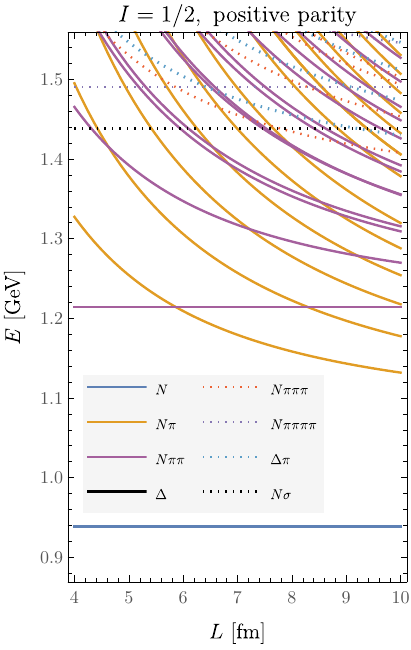}
    \includegraphics[width=0.4\textwidth]{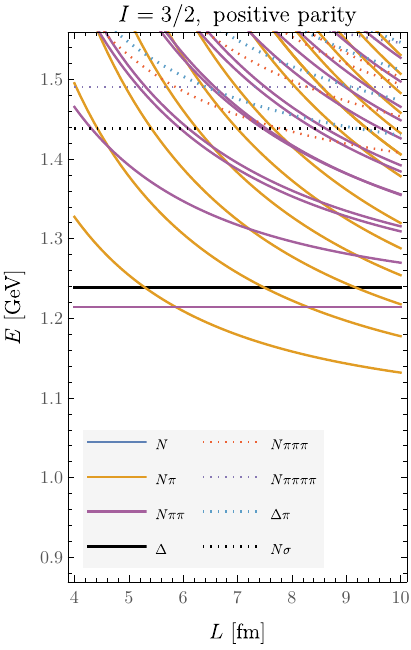}
    \caption{The lowest non-interacting nucleon-meson energy levels for the positive parity $I=1/2$ and $I=3/2$ sectors, where $E_\text{tot} = E_\text{N} + \sum_i E_{\pi, i}$ and $E_N = \sqrt{m_N^2 + \mathbf{p}_N^2}$, $E_{\pi, i} = \sqrt{m_\pi^2 + \mathbf{p}_{\pi, i}^2}$ in a finite box of size $L$.} 
    \label{fig:FV-spectrum}
\end{figure}

In order to compute $N \rightarrow \Delta$ transition form factors with controlled excited-state effects, a necessary first step is to identify interpolating operators that have significant overlap with all QCD energy eigenstates in the vicinity of the $\Delta$ resonance.
This includes multi-hadron states such as $N\pi$ and $N\pi\pi$, and, for a lattice volume with $L \sim 5$--10 fm, several finite-volume $N\pi$ and $N\pi\pi$ states are expected to be relevant based on the non-interacting hadronic effective theory energy spectrum shown in Fig.~\ref{fig:FV-spectrum}.
The same set of $N$, $\Delta$, $N\pi$, and $N\pi\pi$ interpolating operators needed for controlled $N\rightarrow \Delta$ transition matrix element predictions also enables the determination of non-resonant pion neutrinoproduction amplitudes within the same energy range by applying variational methods to correlation-function matrices constructed from these operators~\cite{Fox:1981xz,Michael:1982gb,Luscher:1990ck,Blossier:2009kd}.

\begin{figure}
    \centering
    \includegraphics[width=0.4\textwidth]{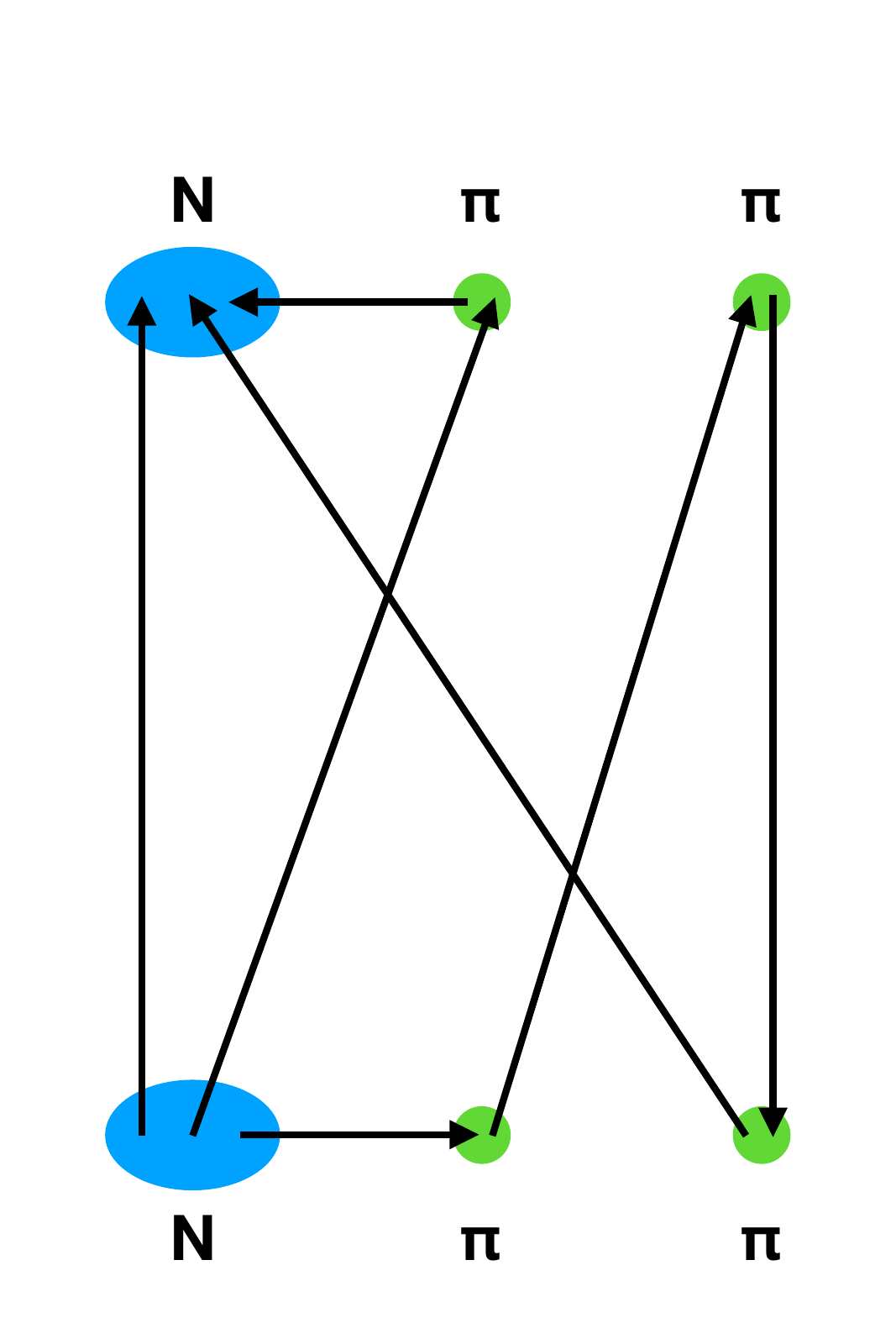}
    \caption{A sample Wick contraction that contributes to the $N\pi\pi$ correlation function.  Note that propagators can originate from either the source or the sink and propagators can begin and end on the same timeslice.  Accessing a range of momentum combinations for the nucleons and pions would na\"ively require prohibitively expensive all-to-all propagators.  This cost can be reduced by sparsening techniques.}
    \label{fig:npipi-contraction}
\end{figure}

Construction of correlation functions requires computing all Wick contractions of propagators between the source and sink interpolating operators.  
This becomes increasingly complicated for operators involving products of multiple hadrons.
For multi-pion systems with maximal isospin, recursive algorithms enable efficient calculations even for very large numbers of pions~\cite{recursive-pions,ryan-pions}.
Only quark propagators connecting the source and sink timeslices are relevant for this particular case, but for generic multi-pion systems as well as nucleon-(multi-)pion systems propagators connecting points on the same timeslice appear as illustrated in Fig.~\ref{fig:npipi-contraction} and require alternative strategies.

For $N \rightarrow N\pi$ correlation functions, a pair of sequential propagator inversions at the nucleon source and pion sink are sufficient~\cite{Barca:2022uhi}.
For $N\pi \rightarrow N\pi$ correlation functions, ``all-to-all'' quark propagators involving source and sink points on all spatial lattice sites and multiple timeslices are needed.
Exact calculations of all-to-all propagators would require cost-prohibitive $O(V^2)$ computational work to solve the Dirac equation for every source point, where $V = (L/a)^3$ is the number of spatial lattice sites.
Approximate all-to-all propagators must therefore be constructed.
This has been achieved in previous works~\cite{Lang:2012db,Kiratidis:2016hda,Lang:2016hnn,Andersen:2017una,Silvi:2021uya,Bulava:2022vpq,Alexandrou:2023elk} using (stochastic variants of) distillation~\cite{distillation, slaph}.
This work explores the application of an alternative all-to-all propagator approximation based on sparsening~\cite{nplqcd-sparsening, xu-feng-sparsening, nplqcd-variational} to $N\pi \rightarrow N\pi$ correlation functions and its extension to $N\pi\pi \rightarrow N\pi\pi$ correlation functions.

Assembling quark propagators into $N\pi\pi \rightarrow N\pi\pi$ correlation functions is, at first glance, even more computationally demanding than constructing all-to-all propagators.
Momentum-projecting all hadrons in $N\pi\pi$ interpolating operators at source and sink separately would require a six-fold volume sum for every Wick contraction, namely
\begin{equation}
    \begin{split}
        C(t) = \sum_{\mathbf{x}_{N_1}} \sum_{\mathbf{x}_{N_2}} \sum_{\mathbf{x}_{\pi_1}} \sum_{\mathbf{x}_{\pi_2}} \sum_{\mathbf{x}_{\pi_3}} \sum_{\mathbf{x}_{\pi_4}} &\left[e^{i\mathbf{x}_{N_1} \cdot \mathbf{p}_{N_1}} e^{i\mathbf{x}_{N_2} \cdot \mathbf{p}_{N_2}} e^{i\mathbf{x}_{\pi_1} \cdot \mathbf{p}_{\pi_1}} e^{i\mathbf{x}_{\pi_2} \cdot \mathbf{p}_{\pi_2}} e^{i\mathbf{x}_{\pi_3} \cdot \mathbf{p}_{\pi_3}} e^{i\mathbf{x}_{\pi_4} \cdot \mathbf{p}_{\pi_4}} \right. \\
        &\left.\langle N(\mathbf{x}_{N_2}, t) \pi(\mathbf{x}_{\pi_3}, t) \pi(\mathbf{x}_{\pi_4}, t) \bar N(\mathbf{x}_{N_1}, 0) \pi^\dagger(\mathbf{x}_{\pi_1}, 0) \pi^\dagger(\mathbf{x}_{\pi_2}, 0) \rangle \right]
    \end{split}
\end{equation}
where spin and flavor indices are omitted for simplicity.  The computational cost of evaluating these volume sums and projecting to operators with definite spin and isospin is $O(V^6)$.

The computational complexity of the contractions can be substantially reduced by factorizing the overall correlation function into sequential propagators through the pions.  Specifically, a sequential propagator through a source pion $\bar \psi \Gamma_\pi \gamma_5 \psi$ (at $t=0$) with momentum $\mathbf{p}$ is given by 
\begin{equation}
  S_{\pi}^{\mathbf{p}} (y|x) = \sum_\mathbf{z} e^{i\mathbf{p} \cdot \mathbf{z}} S(y|\mathbf{z},0) \Gamma_\pi S(\mathbf{z},0 | x),
\end{equation}
where $S(y|x)$ is the quark propagator with source point $x$ and sink point $y$.
Sequential propagators can be computed for all $x$ and $y$ with cost $O(V^3)$ once $S(y|x)$ is known.  
Continuing this process allows one to compute sequential propagators through any ordered sequence of source and sink pions, for a total cost of order $V^3$ times the number of pion momentum combinations.  
For this preliminary analysis, the pion momenta were restricted in order to limit total cost.
A full analysis of all states up to about 1350 MeV --- the $\Delta$ Breit-Wigner mass plus its width \cite{pdg} --- on a lattice with $L \approx 5$ fm would require a 
few hundred momentum combinations: the single nucleon and $\Delta$ states; all $N\pi$ states with $\mathbf{p}_N = \frac{2\pi}{L} \mathbf{n}_N$, $\mathbf{n}_N^2 \leq 2$; and $N\pi\pi$ states with all particles at rest at both source and sink.

These sequential pion propagators can then be assembled into correlation functions by summing over source and sink positions after multiplying by phases corresponding to the nucleon momentum.  At fixed nucleon momentum, this step has a computational cost of $O(V^2)$ times the number of Wick contractions, which can be large (several hundred thousand for $N\pi\pi$).  Converting the resulting correlation-function matrices with different momentum configurations into a basis with definite cubic transformation properties can be performed subsequently with $O(V^0)$ cost.

\subsection{Sparsening}
All three of the steps described above --- all-to-all propagator construction, formation of sequential pion propagators, and contraction of the (sequential) propagators into correlation functions --- would be prohibitively expensive as described above.  To reduce these costs, propagators were computed on a sparse grid of points spaced by $\sparse a$, where $a$ is the lattice spacing and $\sparse$ is the sparsening factor, and were coarsened at the sink by the same factor $\sparse$ \cite{nplqcd-sparsening, xu-feng-sparsening, nplqcd-variational}.  This reduced inversion costs by the ratio of the sparse volume to the full volume $\sparse^3$, sequential pion propagator construction costs by $\sparse^9$, and final contraction costs by $\sparse^6$.  With $\sparse=4$ in this study, this corresponds to a $4^9 \approx 2.6 \times 10^5$-fold reduction in the time to compute sequential propagators.

\begin{figure}
    \centering\includegraphics[width=0.49\textwidth]{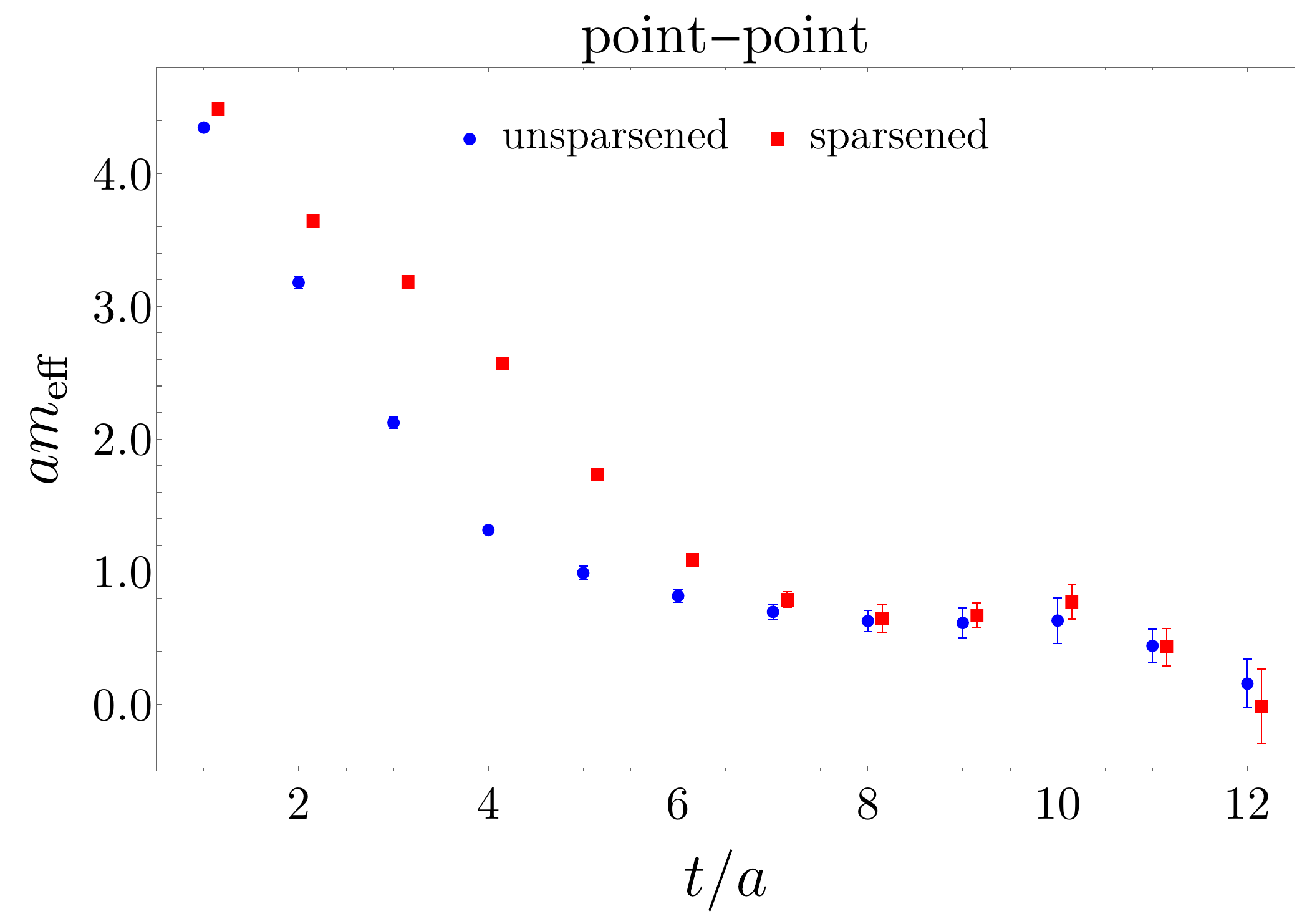}\includegraphics[width=0.49\textwidth]{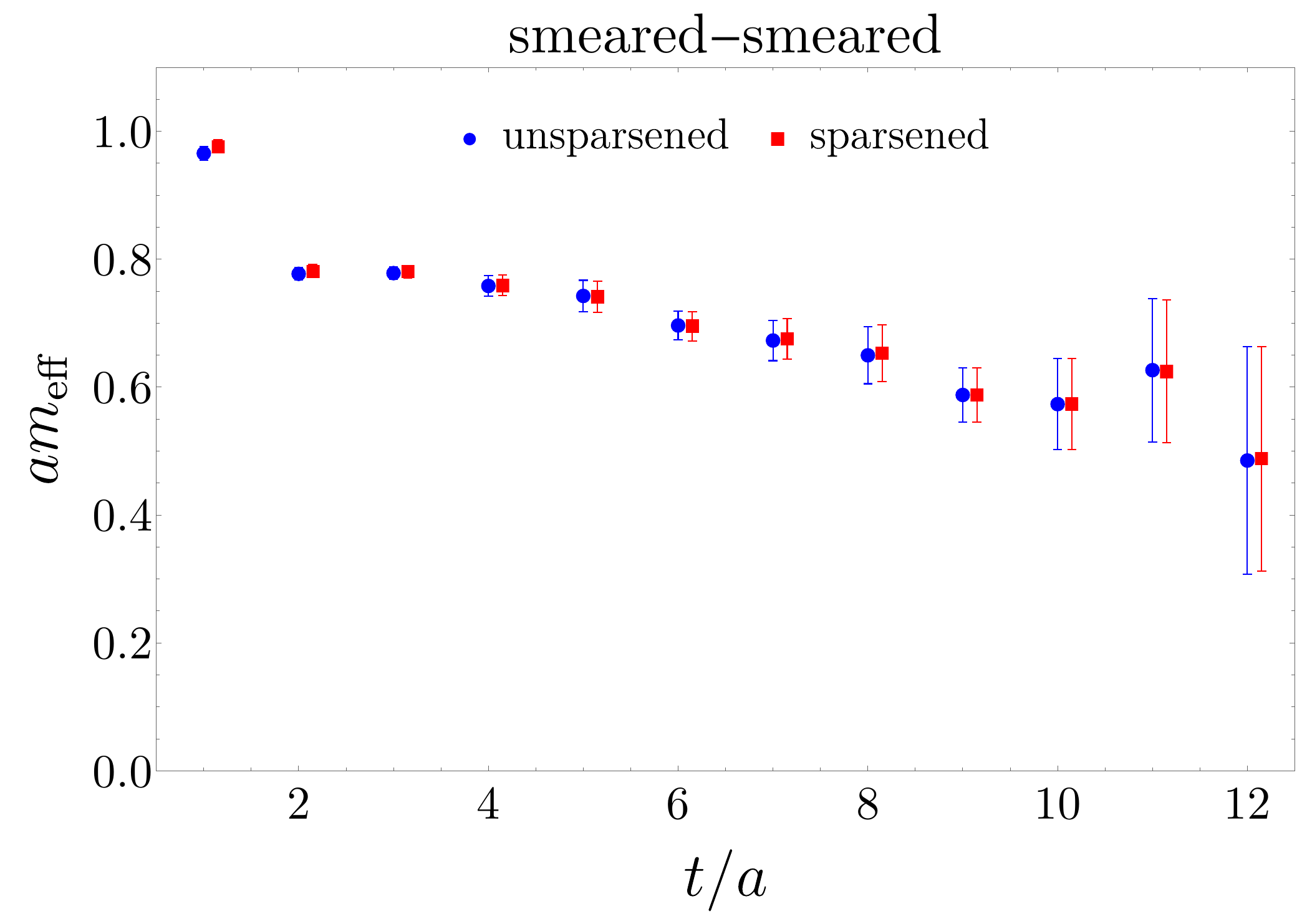}
    \caption{Comparisons of sparsened and unsparsened nucleon effective masses $am_{\text{eff}}(t) = \ln[C(t)/C(t+a)]$ using point-like (left) and Gaussian-smeared (right) interpolating operators. }
    \label{fig:smearing}
\end{figure}

Sparsened correlation functions are not equal to the full correlation functions, even on ensemble average.\footnote{Sparsening with random source and sink locations does lead to correlation functions equal to their unsparsened counterparts on average~\cite{xu-feng-sparsening}; preliminary exploration for this project revealed that this led to significantly larger statistical noise when using point-like sources and sinks.}
In momentum space, constructing a correlation function from sparsened propagators corresponds to incomplete momentum projection.
For momenta below $2\pi/(\sparse a)$, this incomplete momentum projection does not affect the ground state but does introduce additional excited-state contamination from higher-momentum modes.
With point sources and sinks, sparsened nucleon correlation functions differ significantly from their unsparsened counterparts for imaginary times where precise signals are achieved, as shown in Fig.~\ref{fig:smearing}.
With Gaussian smearing~\cite{Gusken:1989ad,Gusken:1989qx} applied to propagator sources and sinks with a smearing radius equal to the sparse lattice spacing ($4a$), the differences between sparsened and unsparsened correlation functions are negligible even for the smallest imaginary times accessible.
This indicates that Gaussian smearing and computationally feasible imaginary-time evolution are sufficient to mitigate the additional excited-state effects introduced by this amount of sparsening.

\subsection{Software Optimizations}
Despite the $\sparse^9$-fold reduction in computational cost from sparsening, sequential propagator construction accounts for a large fraction of the floating-point operations needed to recover the low-lying spectrum.
Since propagators are matrices in spin-color and spacetime, sequential propagator construction corresponds to $(12 V_s) \times (12 V_s)$ matrix multiplication, where $V_s$ is the sparse spatial volume.  Taking $\Gamma_\pi$ to be a parity projector ($P_+ = (1+\gamma_5)/2$ at the source and $P_- = (1-\gamma_5)/2$ at the sink) reduces the matrix dimensions by a factor of 2 and therefore reduces the cost eightfold.

A na\"ive implementation of the matrix multiplication required for sequential propagator construction runs at a small fraction of the theoretical maximum performance of modern processors \cite{leiserson}.  However, given the ubiquity of matrix multiplication in modern computing, there are highly optimized library implementions of BLAS (Basic Linear Algebra Subprograms) for both CPU and GPU architectures.  Linking against Intel MKL BLAS \cite{mkl} provided a 200-fold speedup compared to the na\"ive implementation with nested loops on CPU targets, and linking against NVIDIA cuBLAS \cite{cublas} allowed the sequential propagators to be constructed on GPU targets for even larger speedups.

The remaining pieces of the codebase were also designed to allow GPU offloading.  Propagator inversions on the sparse grid of points were performed the QUDA multigrid inverter optimized for GPUs \cite{quda, quda-mg} and then sparsened and stored to disk.  Contractions of sequential propagators into correlation functions were offloaded to GPUs via the OpenACC framework \cite{openacc}.

\section{Results and Discussion}
As a preliminary demonstration of this method, some of the low-lying states were computed on the $a=0.15$ physical pion mass MILC ensemble, which uses staggered sea quarks \cite{milc}.  Due to the computational and algebraic difficulties of studying baryonic systems with staggered quarks, this study used Wilson-clover valence fermions.  To reduce the impact of exceptional configurations from this mixed-action setup, the original gauge fields were smoothed with gradient flow smearing with flow time $t = 1.0$ \cite{gradient-flow} using the Chroma library \cite{chroma}, and the valence pion mass was set to the slightly heavier-than-physical value of about 170 MeV.

With $L/a = 32, T/a=48$, and a sparsening factor of $\sparse = 4$, quark propagators were computed on an $8^3 \times 48$ grid of 24,576 sources per configuration.  Due to the large number of sources per configuration, only six configurations were needed to obtain almost 150,000 total sources.
This was sufficient to achieve 2\%, 5\%, and 14\% precision on $N$, $N\pi$, and $N\pi\pi$ correlation functions for particles at rest at separations of $t \approx 1.2$ fm, where excited-state effects appear reasonably well controlled.

\begin{figure}
    \centering
    \includegraphics[width=0.49\textwidth]{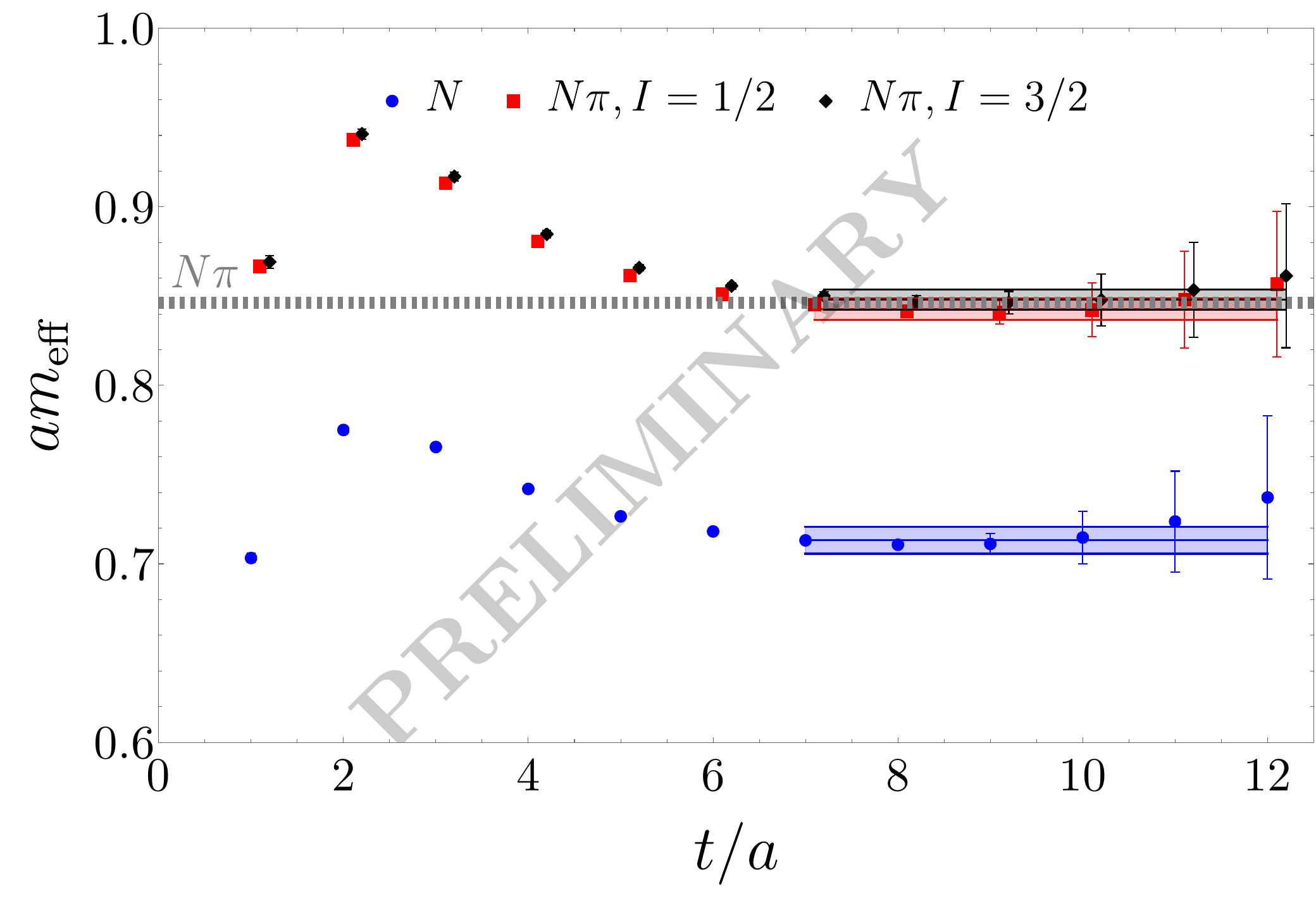}
    \includegraphics[width=0.49\textwidth]{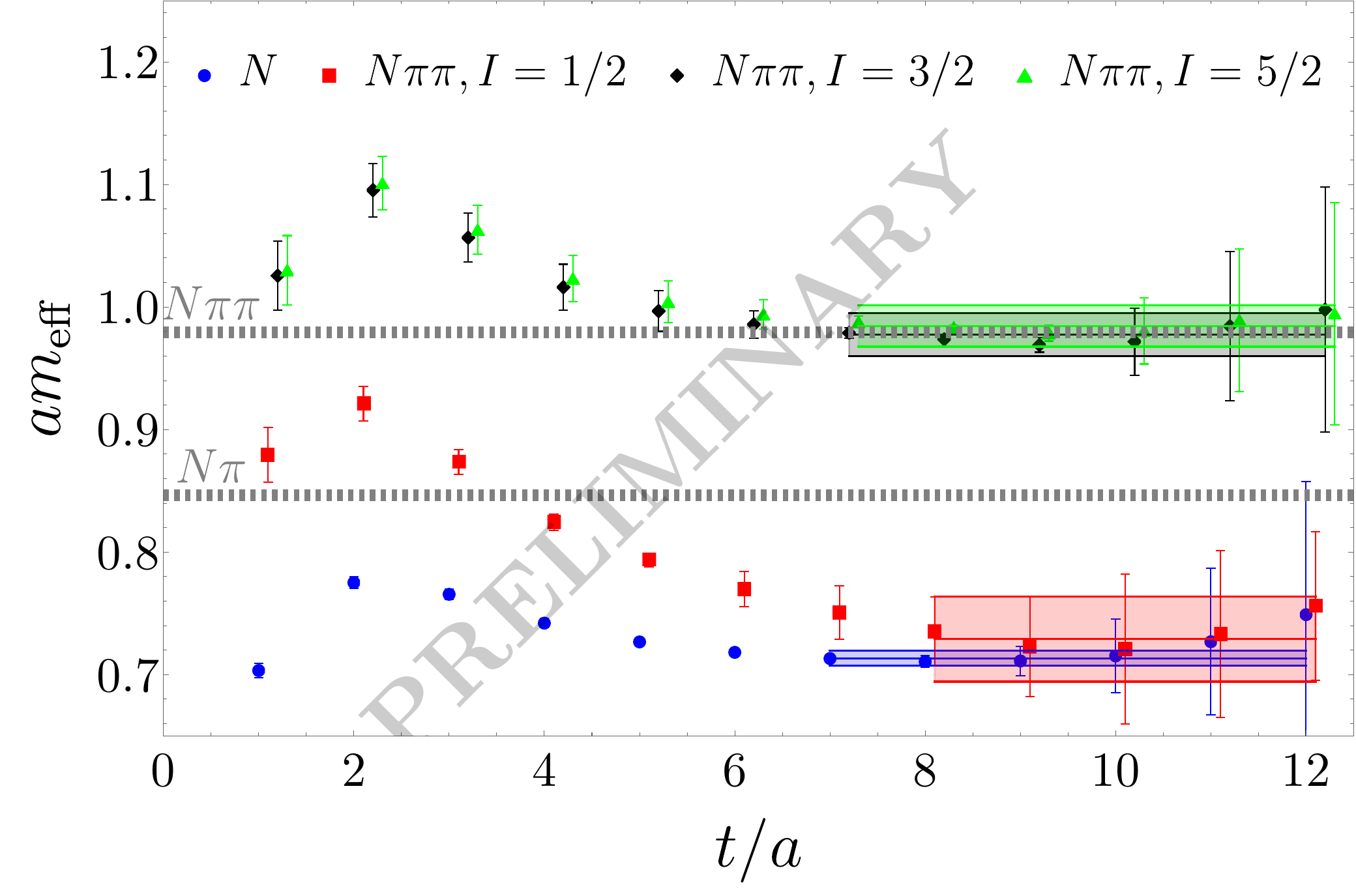}    \includegraphics[width=0.49\textwidth]{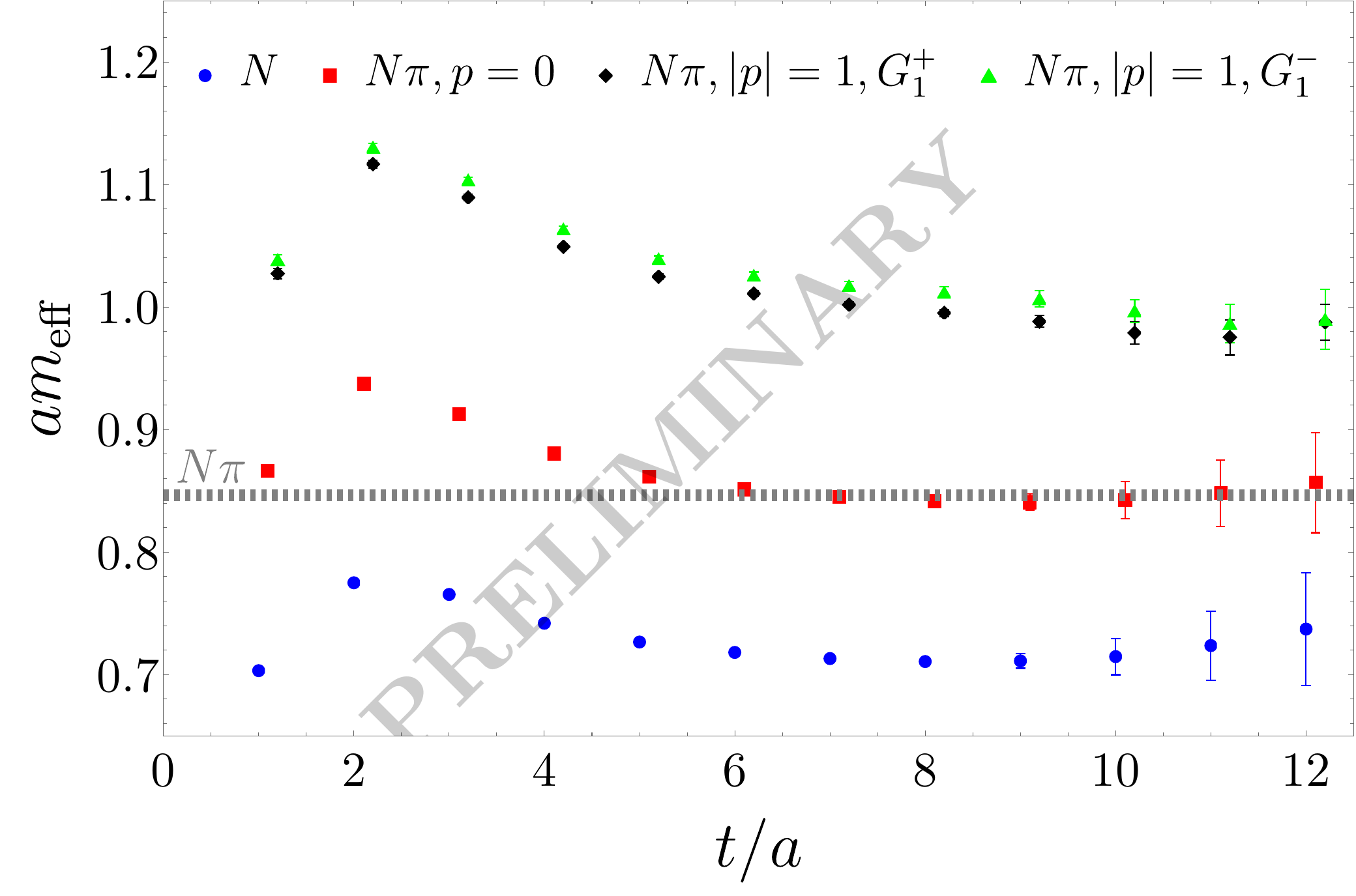}
    \caption{Preliminary effective mass plots of various nucleon-pion states.  (Top left) The single-nucleon effective mass and the two isospin states of $N\pi$.  (Top right) The isospin-$1/2$ channel of $N\pi\pi$ mixes with a single nucleon, while the $I=3/2$ and $I=5/2$ channels are close in energy to the non-interacting $N\pi\pi$ energy level.  (Bottom) The two parity sectors of the $I=1/2$ $N\pi$ interpolating operator have effective masses at accessible Euclidean time exceeding those of the single nucleon or $N\pi$ at rest.}
    \label{fig:results}
\end{figure}

A single nucleon and $N\pi$ at rest form the ground states of their respective sectors ($I=1/2$, positive parity; $I=1/2$, negative parity; and $I=3/2$, negative parity).  The two isospin channels of $N\pi$ both lie very close to the non-interacting $N\pi$ threshold, although when accounting for correlations between configurations, the repulsive $I=3/2$ channel lies at slightly higher energy than the attractive $I=1/2$ channel.

In $N\pi\pi$ systems, where the overall parity is positive, the picture is more interesting.  While in the $I=5/2$ channel, $N\pi\pi$ forms the ground state, this is no longer the case in the $I=1/2$ channel, where $N$ and $N\pi\pi$ have the same quantum numbers.  As such, an interpolating operator designed to produce $N\pi\pi$ with $I=1/2$ will, in general, contain an admixture of the single-nucleon ground state and will therefore plateau to $m_N$ at sufficiently large Euclidean time.  With the interpolating operators considered in this work, this overlap was sufficiently large to see this plateau even at the moderate Euclidean time accessible here.  Disentangling the two physical states corresponding to $N$ and $N\pi\pi$ with these quantum numbers will require solving the generalized eigenvalue problem (GEVP) \cite{gevp} to isolate the two energy levels.

With the addition of relative momenta, $N\pi$ systems can have either positive or negative parity, depending on the angular momentum of the $N\pi$ system.  In the $I=1/2$ channel, the positive and negative parity systems have overlap onto $N$ and $N\pi$ at rest, respectively, so correlation functions sourced by these interpolating operators should ultimately plateau to $m_N$ and $E_{N\pi}$, respectively.  While the negative parity correlation function is higher energy than the positive parity version, neither has plateaued to its true ground state, indicating relatively poor overlap between boosted $N\pi$ systems and either $N$ or $N\pi$ at rest.  A careful GEVP analysis including single-nucleon, $N\pi$, and $N\pi\pi$ interpolating operators will be necessary to isolate the full tower of states in each of the parity sectors.

\section{Conclusion}
This analysis only considered nucleon-pion states up to about $1250~\text{MeV} \approx m_\Delta \approx m_N + 2m_\pi \approx E_N(|\mathbf{n}| = 1) + E_\pi(|\mathbf{n}| = 1)$.  A more thorough calculation would include states of somewhat higher energy by increasing the momenta of various particles, as well as including $\Delta$ and $N\sigma$ interpolating operators that were not considered here.  Approaching the $N\pi\pi\pi$ threshold in the positive parity sector in an $L \approx 5$ fm box would require $N\pi$ states with up to 2 units of lattice momentum and $N\pi\pi$ with the nucleon carrying up to $\sqrt{2}$ units of momentum.  Such a study would be computationally intensive due to the large number of such momentum combinations at both the source and sink but possible with the codebase developed here, and it would allow a more careful study of the $\Delta$ resonance and a more complete variational analysis of the low-lying single-baryon sector.  Repeating the analysis at finer lattice spacings in future work would allow discretization artifacts to be analyzed.

More long-term, neutrino experiments will require $N \rightarrow \Delta$ axial form factors.  Since the $\Delta$ is unstable, this will require studying $N \rightarrow N\pi, N\pi\pi$ transitions in the energy regime around $m_\Delta$.  Understanding the spectrum and the resultant excited state contamination will be necessary to construct good interpolating operators for the various states in this energy regime and therefore predict the form factors needed to describe neutrino-nucleus scattering in the resonance region with good control over excited states.

\section{Acknowledgements}

The authors would like to thank Raul Brice{\~n}o, William Detmold, William Jay, Gurtej Kanwar, Phiala Shanahan, Jim Simone, Ruth Van de Water, and Andr\'e Walker-Loud for useful discussions.
This manuscript has been authored by Fermi Research Alliance, LLC under Contract No.~DE-AC02-07CH11359 with the U.S. Department of Energy, Office of Science, Office of High Energy Physics.

\bibliographystyle{JHEP}
\bibliography{refs}

\end{document}